\documentclass[prl,twocolumn,superscriptaddress,showkeys,showpacs]{revtex4}
\usepackage{epsf}

\def\llangle{\langle\!\langle}
\def\rrangle{\rangle\!\rangle}

\begin{document}

\title{Quantum Trajectory Distribution for Weak Measurement\\
       of a Superconducting Qubit: Experiment meets Theory}
\author{Parveen Kumar}
\email{parveenkumar@iisc.ac.in}
\affiliation{Centre for High Energy Physics,
             Indian Institute of Science, Bangalore 560012, India}
\author{Suman Kundu}
\email{suman.kundu@tifr.res.in}
\affiliation{Department of Condensed Matter Physics and Materials Science,
             Tata Institue of Fundamental Research, Mumbai 400005, India}
\author{Madhavi Chand}
\email{chand@tifr.res.in}
\affiliation{Department of Condensed Matter Physics and Materials Science,
             Tata Institue of Fundamental Research, Mumbai 400005, India}
\author{R. Vijayaraghavan}
\email{r.vijay@tifr.res.in}
\affiliation{Department of Condensed Matter Physics and Materials Science,
             Tata Institue of Fundamental Research, Mumbai 400005, India}
\author{Apoorva Patel}
\email{adpatel@iisc.ac.in}
\affiliation{Centre for High Energy Physics,
             Indian Institute of Science, Bangalore 560012, India}

\date{\today}

\begin{abstract}
Quantum measurements are described as instantaneous projections in textbooks.
They can be stretched out in time using weak measurements, whereby one can
observe the evolution of a quantum state as it heads towards one of the
eigenstates of the measured operator. This evolution can be understood as a
continuous nonlinear stochastic process, generating an ensemble of quantum
trajectories, consisting of noisy fluctuations on top of geodesics that
attract the quantum state towards the measured operator eigenstates.
The rate of evolution is specific to each system-apparatus pair, and
the Born rule constraint requires the magnitudes of the noise and the
attraction to be precisely related. We experimentally observe the entire
quantum trajectory distribution for weak measurements of a superconducting
qubit in circuit QED architecture, quantify it, and demonstrate that it
agrees very well with the predictions of a single-parameter white-noise
stochastic process. This characterisation of quantum trajectories is a
powerful clue to unraveling the dynamics of quantum measurement, beyond
the conventional axiomatic quantum theory.
\end{abstract}



\maketitle

\noindent{\bf Motivation:}
The textbook formulation of quantum mechanics describes measurement
according to the von Neumann projection postulate, which states that
one of the eigenvalues of the measured observable is the measurement
outcome and the post-measurement state is the corresponding eigenvector.
With $P_i$ denoting the projection operator for the eigenstate $|i\rangle$,
\begin{eqnarray}
|\psi\rangle \longrightarrow P_i |\psi\rangle / |P_i |\psi\rangle| , &&\\
P_i = P_i^\dagger ,~~ P_i P_j = P_i \delta_{ij} ,~~ && \sum_i P_i = I .
\end{eqnarray}
This change is sudden, irreversible, consistent on repetition, and
probabilistic in the choice of ``$i$". Which ``$i$" would occur in a
particular experimental run, is not specified; only the probabilities
of various outcomes are specified, requiring an ensemble interpretation
for the outcomes. These probabilities follow the Born rule,
\begin{equation}
\label{Bornrule}
prob(i) = \langle\psi|P_i|\psi\rangle = Tr(P_i\rho) ~,~~
\rho \longrightarrow \sum_i P_i \rho P_i ~,
\end{equation}
and the ensemble evolution takes initially pure states to mixed states.

Over the years, many attempts have been made to unravel the dynamics of
this process \cite{Wheeler,Braginsky}. The framework of environmental
decoherence is an important step that provides continuous interpolation of
the sudden projection. In this framework, both the system and its environment
(which includes the measuring apparatus) follow the unitary Schr\"odinger
evolution, and the unobserved degrees of freedom of the environment are
``summed over" to determine how the remaining observed degrees of freedom
evolve. The result is still an ensemble description, but it provides a
quantitative understanding of how the off-diagonal elements of the reduced
density matrix $\rho$ decay \cite{Zeh,Wiseman}. Subsequently, solution of
the ``measurement problem" requires decomposing the ensemble into individual
quantum contributions (i.e. which ``$i$" will occur in which experimental
run), and that forces us to look beyond the closed unitary Schr\"odinger
evolution.

Some of the attempted decompositions of the quantum ensemble are physical,
e.g. introduction of hidden variables with novel dynamics or ignored
interactions with known dynamics \cite{Bohm,Ghirardirev,Penrose,Bassirev}.
Some other attempts philosophically question what is real and what is
observable, in principle as well as by human beings with limited capacity
\cite{Everett,GMH,QBism}. Although these attempts are not theoretically
inconsistent, none of them have been positively verified by
experiments---either they are untestable or only bounds exist
on their parameters \cite{Bassirev,Bassi}.

We focus here on the extension of quantum mechanics that describes the
measurement as a continuous nonlinear stochastic process. It is a particular
case of the class of stochastic collapse models that add a measurement
driving term and a random noise term to the Schr\"odinger evolution
\cite{Bassirev}. We look at these terms in an effective theory approach,
without assuming a specific collapse basis (e.g. energy or position basis)
or a specific collapse interaction (e.g. gravity or some other universal
interaction). In this expanded view, the collapse process can be specific
to each system-apparatus pair and need not be universal. Such a setting is
necessary to understand the quantum state evolution during continuous
measurements of superconducting transmon qubits \cite{VijayRev}, where the
collapse basis as well as the system-apparatus interaction strength can be
varied by changing the control parameters and without changing the apparatus
mass or size or position.

Realisation of quantum measurement as a continuous stochastic process is
tightly constrained by the well-established properties of quantum dynamics
\cite{Pearle,Gisin,Diosi}. A precise combination of the attraction towards
the eigenstates and unbiased noise is needed to reproduce the Born rule
as a constant of evolution \cite{Gisin}. Two of us have emphasised recently
that this is a fluctuation-dissipation relation \cite{BornRule}. It points
to a common origin for the stochastic and the deterministic contributions
to the measurement evolution, analogous to both diffusion and viscous damping
arising from the same underlying molecular scattering in statistical physics.
Moreover, for a binary measurement (i.e. when the measured operator has only
two eigenvalues), the complete quantum trajectory distribution is predicted
in terms of only a single dimensionless evolution parameter.

In this work, we experimentally observe quantum trajectories for
superconducting qubits, using weak measurements \cite{Bub} that stretch
out the evolution time from the initial state to the final projected state.
Going beyond previous experiments \cite{Murch,Weber} that deduced the most
likely evolution paths from the observed trajectories, we quantitatively
compare the entire observed trajectory distribution with the single parameter
theoretical prediction, to test the validity of the nonlinear stochastic
evolution model. We also observe the quantum trajectories for time scales
going up to the relaxation time; the relaxation substantially alters the
evolution, and we show that the relaxation effects can be successfully
described by a simple modification of the theoretical model.

\noindent{\bf Theoretical Predictions:}
We consider evolution of a qubit undergoing binary weak measurement in
absence of any driving Hamiltonian, and explicitly include the effect of
a finite excited state relaxation time $T_1$. In the continuous stochastic
quantum measurement model, the evolution depends on the nature of the noise;
we consider the particular case of white noise that is appropriate for weak
measurements of transmons \cite{Korotkov}. In this quantum diffusion
scenario, with $|0\rangle$ and $|1\rangle$ as the measurement eigenstates,
the density matrix evolves according to \cite{qubitevol}:
\begin{equation}
\label{Strat1qd}
{d\over dt}\rho_{00} = 2g(w_0-w_1)~\rho_{00}~\rho_{11}
                     + {1\over T_1}\rho_{11} ~.
\end{equation}
Here $g(t)$ is the system-apparatus coupling, and $w_i(t)$ are real weights
representing evolution towards the two eigenstates. The weights satisfy
$w_0+w_1=1$, and
\begin{equation}
\label{oneqweight}
w_0 - w_1 = \rho_{00} - \rho_{11} + \sqrt{S_\xi}~\xi ~,
\end{equation}
where the unbiased white noise with spectral density $S_\xi$ obeys
$\llangle\xi(t)\rrangle=0$ and $\llangle\xi(t)\xi(t')\rrangle = \delta(t-t')$.

This evolution is a stochastic differential process on the interval $[0,1]$,
with perfectly absorbing boundaries. The Born rule becomes a constant of
evolution when $gS_\xi=1$ \cite{Pearle,Gisin}, which we impose henceforth.
The It\^o form is convenient for numerical simulations of quantum trajectories:
\begin{equation}
\label{Ito1qd}
d\rho_{00} = 2\sqrt{g}~\rho_{00}~\rho_{11}~dW + (1/T_1)\rho_{11}~dt ~,
\end{equation}
where the Wiener increment satisfies $\llangle dW(t)\rrangle=0$ and
$(dW(t))^2=dt$. Although we are unable to integrate Eq.(\ref{Ito1qd})
exactly, exact integrals of each of the two terms on its right-hand-side
are known. We simulate quantum trajectories using a Gaussian distribution
for $dW$ and a symmetric Trotter-type integration scheme. The discretisation
error is then $O((dt)^2)$ for individual steps, and is made negligible by
making $dt$ sufficiently small.

The probability distribution of the quantum trajectories, $p(\rho_{00},t)$,
obeys the Fokker-Planck equation:
\begin{eqnarray}
{\partial p(\rho_{00},t) \over \partial t}
&=& 2g {\partial^2 \over \partial^2 \rho_{00}}
    \Big( \rho_{00}^2 \rho_{11}^2 p(\rho_{00},t) \Big) \nonumber\\
&-& {1\over T_1} {\partial \over \partial \rho_{00}}
    \Big( \rho_{11} p(\rho_{00},t) \Big) ~.
\label{FPevol}
\end{eqnarray}
Its solution is easier to visualise after the map
\begin{equation}
\tanh(z) = 2\rho_{00}-1 = \rho_{00}-\rho_{11} ~,
\end{equation}
from $\rho_{00}\in[0,1]$ to $z\in[-\infty,\infty]$.
With the initial condition $p(\rho_{00},0)=\delta(x)$, and in absence of
any relaxation, the exact solution consists of two non-interfering parts
with areas $x$ and $1-x$, monotonically travelling to the boundaries at
$\rho_{00}=1$ and $0$ respectively \cite{Pearle,Gisin,pathintdist}.
In terms of the variable $z$, these parts of $p(z,t)$ are Gaussians, with
centres at $z_\pm(t) = \tanh^{-1}(2x-1) \pm gt$ and common variance $gt$.
Excited state relaxation introduces an additional drift in the evolution,
which makes the part heading to $\rho_{00}=1$ grow at the expense of the
part heading to $\rho_{00}=0$.

Other than the unavoidable relaxation time $T_1$, the entire quantum
trajectory distribution is determined in terms of the single dimensionless
evolution parameter $\tau(g,t)\equiv\int_0^t g(t')dt'$. A strong measurement,
$\tau>10$ \cite{strongtau}, is essentially a projective measurement. In weak
measurement experiments on transmons, the coupling $g$ is a tunable parameter,
and the intervening stages between the initial state and the final projective
outcome can be observed as $\tau$ gradually increases. We next present the
results of such an experiment.

\noindent{\bf Experimental Results:}
Our experiments were carried out on superconducting 3D transmon qubits,
placed inside a microwave resonator cavity and dispersively coupled to it.
The transmon is a nonlinear oscillator \cite{transmonref}, consisting of
a Josephson junction shunted by a capacitor; the two lowest quantum levels
are treated as a qubit, which possesses good coherence and is insensitive
to charge noise. Measurements of the qubit are performed in the circuit
QED architecture \cite{VijayRev}, by probing the cavity with a resonant
microwave pulse; the amplitude of the microwave pulse controls the
measurement strength. The scattered wave is amplified
by a near-quantum-limited Josephson parametric amplifier \cite{JPA}
operated in the phase-sensitive mode, so that only the quadrature
containing information about the $\sigma_z$ component of the qubit is
amplified \cite{VijayRev}. The amplified single-quadrature signal is
extracted as a measurement current using standard homodyne detection.
We provide more details of the device and the experimental setup in the
Supplementary Information.

\begin{figure}[b]
%{\epsfxsize=8.6truecm \epsfbox{eigen_dist_Wk015_03_halfus.eps}}
{\epsfxsize=8.6truecm \epsfbox{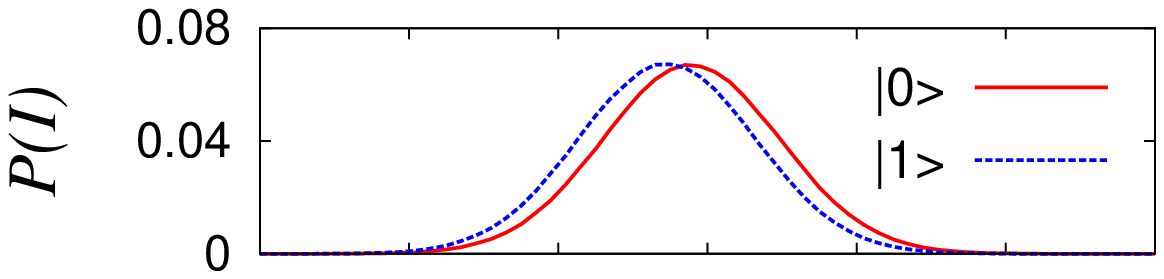}}

\vspace{-4mm}

{\epsfxsize=8.6truecm \epsfbox{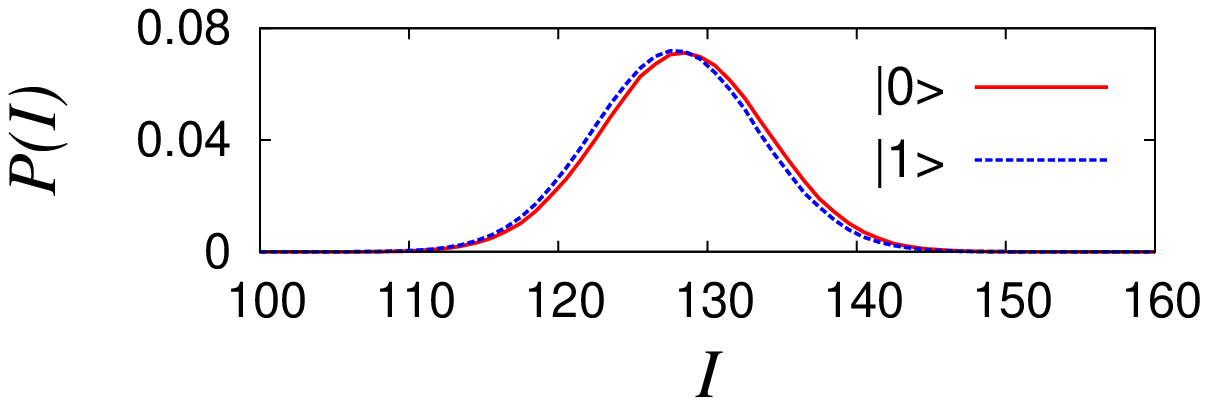}}
\caption{Measurement current distributions for the qubit eigenstates after
evolution for $0.5\mu s$, for an ensemble of $10^6$ trajectories. They are
Gaussians to high accuracy. The bottom figure corresponds to a weaker
system-apparatus coupling than the top one. The system-apparatus coupling
increases mostly by an increase in $\Delta I$ without much change in
$\sigma$. Gaussian fits give the parameters:
(bottom) $I_0=128.443(2)$, $I_1=127.856(2)$, $\sigma=5.56(3)$, and
(top) $I_0=128.919(2)$, $I_1=127.286(2)$, $\sigma=5.93(2)$.}
% (bottom) $\sigma_0=5.589(2)$, $\sigma_1=5.536(2)$.
% (top) $\sigma_0=5.955(2)$, $\sigma_1=5.913(2)$.
\label{eigendist}
\end{figure}

The qubit eigenstates produce Gaussian distributions for the measurement
current, centred at $I_0$ and $I_1$, and with variance $\sigma^2$
\cite{spline}. The measurements are weak when $\Delta I =|I_0-I_1|\ll\sigma$,
as illustrated in Fig.~\ref{eigendist}.
The integrated current measurement gives the quantum trajectory evolution,
according to the Bayesian prescription \cite{Korotkov} (note that
$\rho_{11}=1-\rho_{00}$):
\begin{eqnarray}
\label{Bayesevol1}
{\rho_{00}(t+\Delta t) \over \rho_{11}(t+\Delta t)} &=&
{\rho_{00}(t) \over \rho_{11}(t)} {\exp[-(I_m(\Delta t)-I_0)^2/2\sigma^2]
\over \exp[-(I_m(\Delta t)-I_1)^2/2\sigma^2]} ~,\\
I_m(\Delta t) &=& {1 \over \Delta t} \int_0^{\Delta t} I(t')~dt' ~.
\end{eqnarray}
Simultaneous relaxation of the excited state produces:
\begin{equation}
\label{Bayesevol2}
\rho_{11}(t+\Delta t) = \rho_{11}(t) \exp(-\Delta t/T_1) ~.
\end{equation}
We construct complete quantum trajectories by combining these two evolutions
in a symmetric Trotter-type scheme. This construction has been shown to be
fully consistent with direct quantum state tomography at any time $t$
\cite{Murch}; even though the measurement extracts only partial information,
its back-action on the qubit is completely known, and the qubit evolution from
a known starting state can be precisely constructed \cite{Hatridge,Murch}.

We prepared the qubit ground state by relaxation for $500\mu s$, followed
by a heralding strong measurement \cite{johnson}. After a $3\mu s$ delay
to empty the measurement cavity of photons, the initial state in the
XZ-subspace of the Bloch sphere was created by an excitation pulse at the
qubit transition frequency. The duration of the pulse was fixed by demanding
that an immediate strong measurement gives outcomes $|0\rangle$ and
$|1\rangle$ with the desired probabilities. Following the state preparation,
weak measurements were performed to obtain $40\mu s$ long trajectories with
time step $\Delta t=0.5\mu s$. The process was repeated to generate an
ensemble of $10^6$ trajectories.

We determined $T_1$ from the decay rate of the ensemble averaged weak
measurement current, after initialising the qubit in the excited state.
We observed that it depended on the control parameters that fixed the
system-apparatus coupling, i.e. the amplitude of the cavity drive.
Hence for each system-apparatus coupling, we extracted a separate $T_1$
and used that in Eqs.(\ref{Strat1qd},\ref{FPevol})
to analyse the quantum trajectories. Our strong measurements were
performed over a single time step $\Delta t_s=0.5\mu s$, so we estimated
$(1-e^{-\Delta t_s/T_1})$ as the uncertainty in the initial state.
The parameters $I_0,I_1,\sigma$ were extracted by making Gaussian fits
to the current probability distributions for the ground and the excited
states for each measurement strength (see Fig.~\ref{eigendist}). We found
that the statistical errors in these parameters are small compared to their
systematic errors arising from their variations over time. To estimate the
systematic errors, we monitored variations in the parameters
$T_1,I_0,I_1,\sigma$ for a duration of 3 hours. (It took about 10 minutes
to generate the trajectory ensemble for each choice of the system-apparatus
coupling and the initial state.) Then we created different trajectory
distributions from the experimental current data, by varying the initial
state, $T_1,I_0,I_1$, one at a time within their range of fluctuations, and
added the shifts in the trajectory distributions in quadrature to estimate
the total systematic error. (More details are given in the Supplementary
Information.) We did not worry about variations in $\sigma$, because they
get absorbed in the value of the fit parameter $\tau(g,t)$ \cite{ampeff}.
Overall, the dominant sources of error were the uncertainties in the initial
state, $I_0$ and $I_1$.

\begin{figure*}[t]
{\epsfxsize=5.8truecm \epsfbox{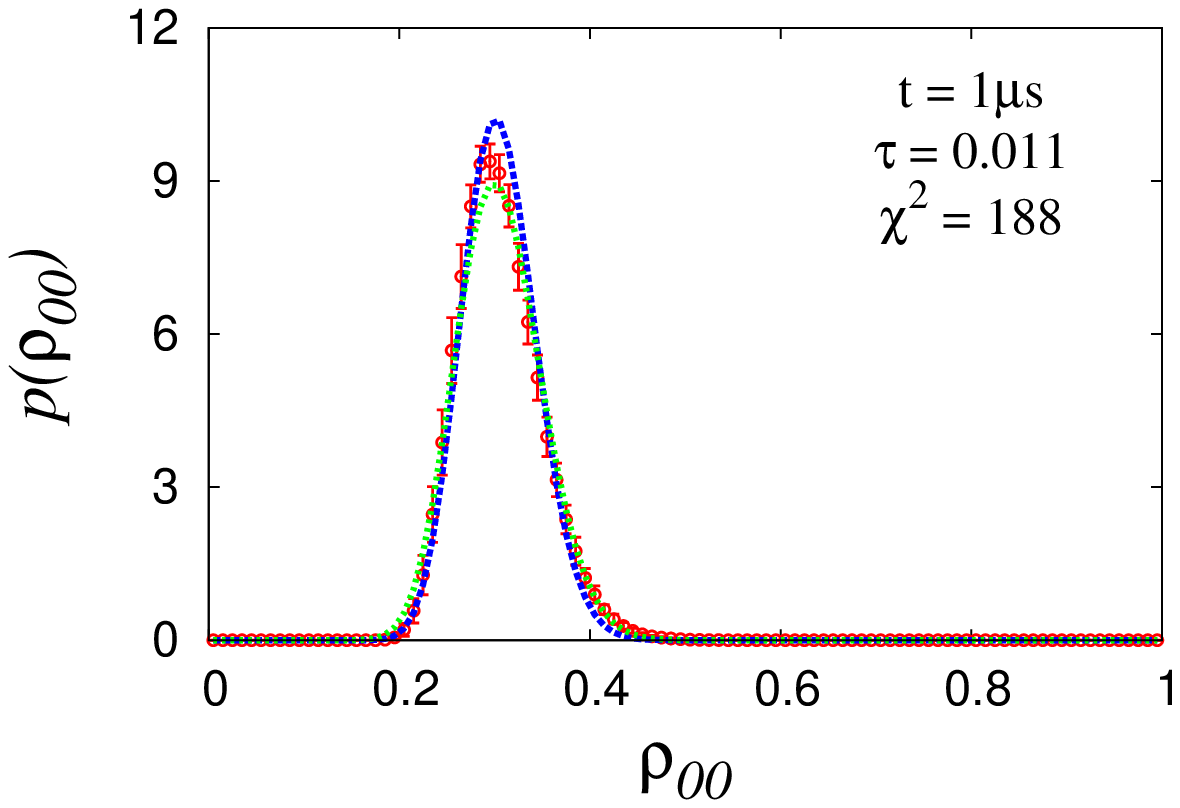}}
{\epsfxsize=5.8truecm \epsfbox{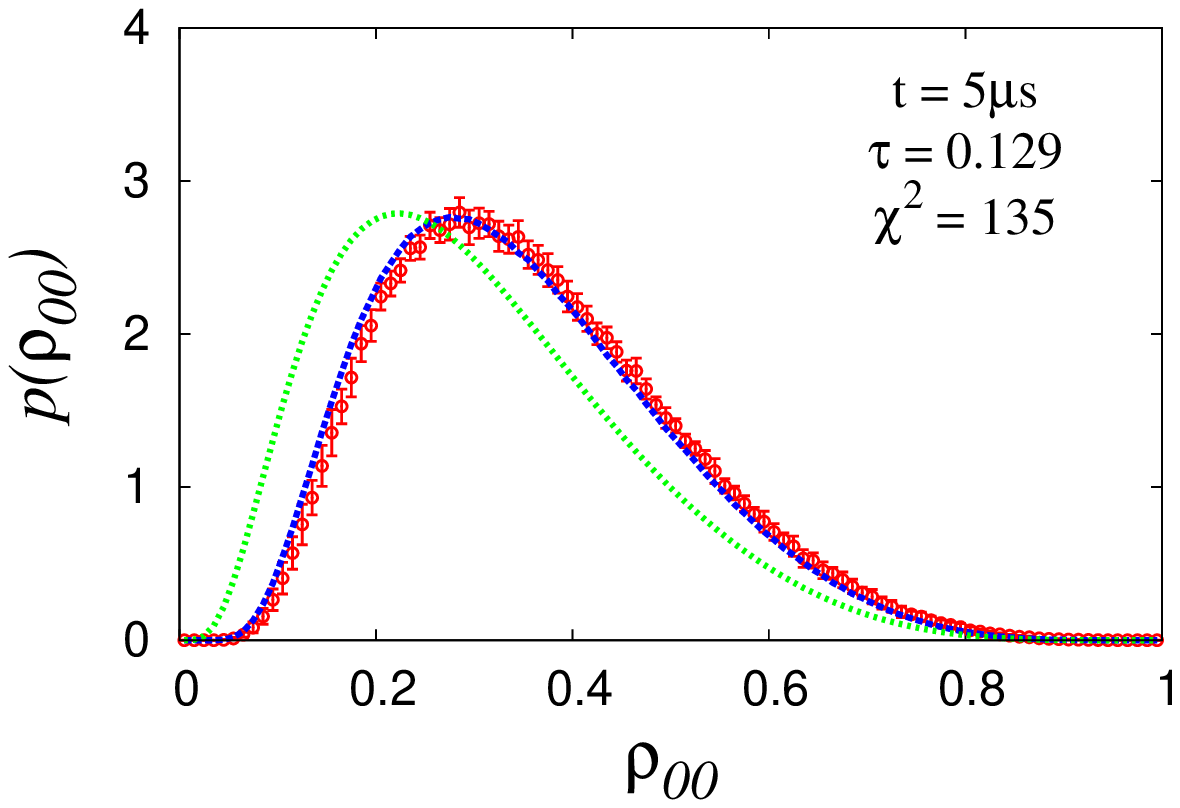}}
{\epsfxsize=5.8truecm \epsfbox{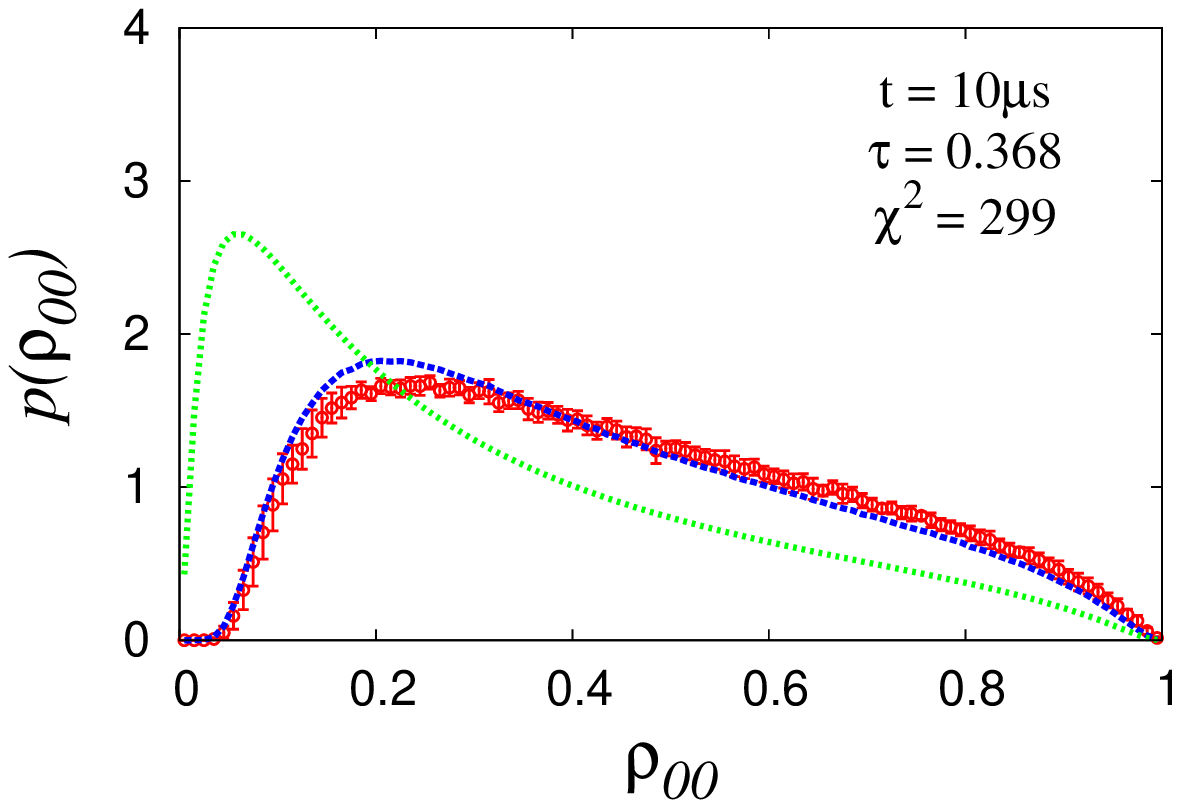}}
%\tau=0.011,0.129,0.368

{\epsfxsize=5.8truecm \epsfbox{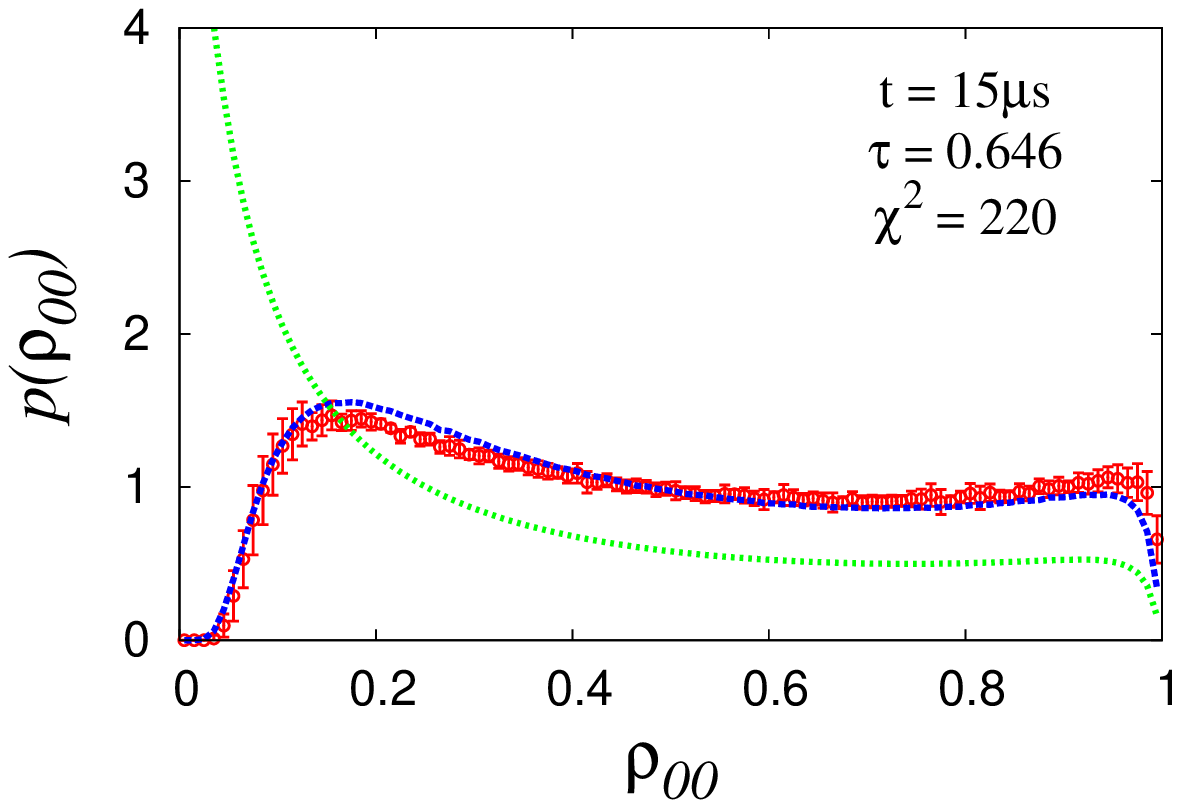}}
{\epsfxsize=5.8truecm \epsfbox{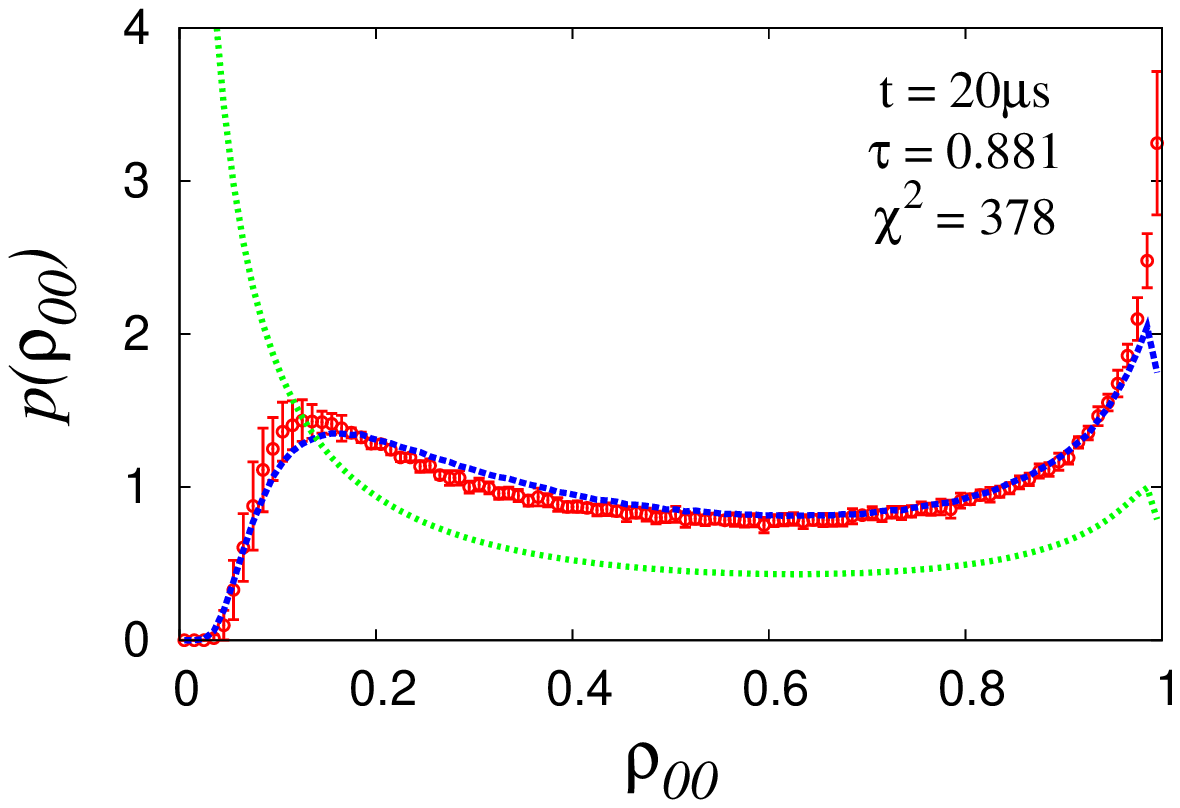}}
{\epsfxsize=5.8truecm \epsfbox{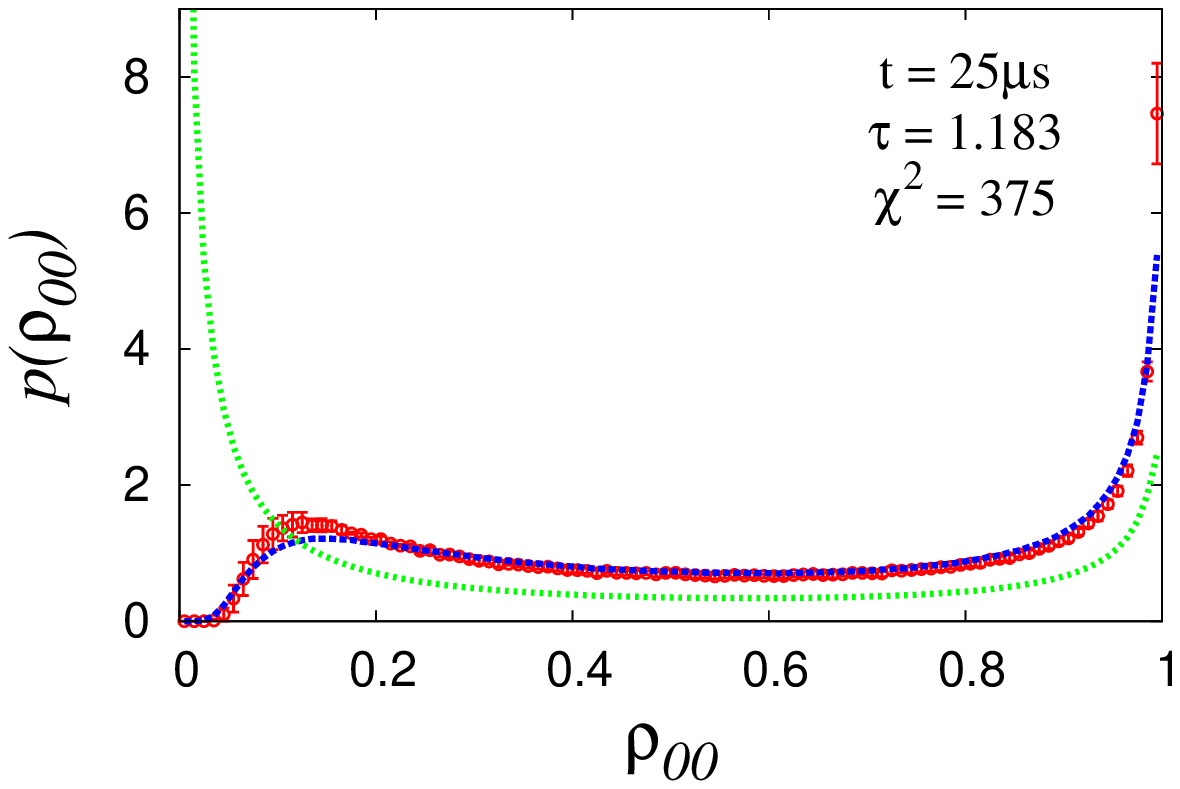}}
%\tau=0.646,0.881,1.183

\caption{Evolution of the quantum trajectory distribution for the weak
Z-measurement of a transmon, with the initial state $\rho_{00}=0.305(3)$.
The histograms with bin width 0.01 (red) represent the experimental data
for an ensemble of $10^6$ trajectories. The trajectory parameters (with
errors) were $T_1=45(4)\mu s$, $\Delta t=0.5\mu s$, $I_0=128.44(2)$,
$I_1=127.68(3)$, $\sigma=5.50(1)$. The blue curves are the best fits to
the quantum diffusion model distribution including relaxation, with the
evolution parameter $\tau \in [0,1.2]$; the green curves show the
theoretical distributions with the same evolution parameter but with
$T_1$ set to infinity.}
\label{trajdist}
\end{figure*}

\begin{figure}[b]
{\epsfxsize=8.6truecm \epsfbox{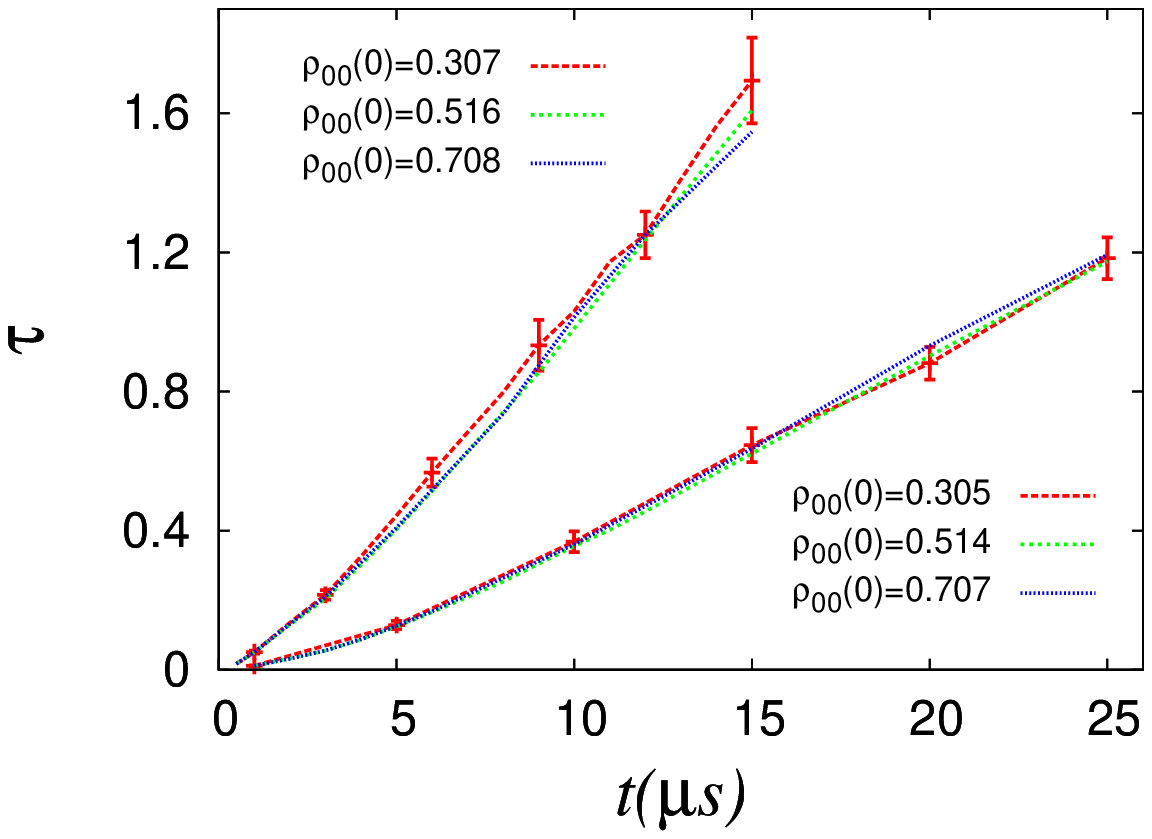}}
\caption{The best fit values of the time integrated measurement coupling
$\tau$ for two values of the system-apparatus coupling, when experimental
data for weak Z-measurement of a transmon with different initial states
$\rho_{00}(0)$ are compared to the theoretical predictions. The stronger
coupling (top curves) had $T_1=25(3)\mu s$, while the weaker coupling
(bottom curves) had $T_1=45(4)\mu s$. It is obvious that $\tau$ is essentially
independent of the initial state, and varies almost linearly with time after
a slower initial build-up. The error bars (only some are shown for the sake of
clarity) correspond to changes in $\tau$ that would change the $\chi^2$ values
for the trajectory distribution fits by 100.}
\label{gtvst}
\end{figure}

Our observed evolution of the quantum trajectory distribution, with the
data divided into 100 histogram bins, is shown in Fig.~\ref{trajdist}, for
a particular choice of the system-apparatus coupling and the initial qubit
state. We have compared it with the simulated distribution of $10^7$
trajectories, obtained by integrating Eq.(\ref{Ito1qd}) with $\tau(g,t)$
as the only fit parameter. (We generated the simulated distributions for
many values of $\tau$ to find the best fit, and we cross-checked that the
simulated distributions agree with the solution of Eq.(\ref{FPevol})
obtained using a symmetric Trotter-type integration scheme.) This fitting
of the entire trajectory distribution goes much beyond looking at just the
mean and the variance of the distribution. With 100 data points and only
one fit parameter, $\chi^2$ values less than a few hundred indicate good
fits, and that is what we find. Also shown in Fig.~\ref{trajdist} are the
theoretically calculated trajectory distributions in absence of any
relaxation, i.e. $T_1\rightarrow\infty$. We see that relaxation alters
the distributions substantially, even for $t=0.1T_1$, and the quantum
diffusion model of Eq.(\ref{Strat1qd}) successfully accounts for the
changes.

We point out that the mismatch between theory and experiment grows with
increasing evolution time, quite likely due to magnification of small initial
uncertainties due to the iterative evolution. On closer inspection, we also
observe certain systematic discrepancies, which are likely due to experimental
imperfections that we have not accounted for. They include transients caused
by photons left in the cavity after the initial heralding pulse, and
contamination due to occupation of the higher excited states of the transmon
\cite{drift}. More stringent tests of the quantum diffusion model would need
to control and account for them.

We have plotted the best fit $\tau$ values against $t$ in Fig.~\ref{gtvst},
for two values of the system-apparatus coupling and for different qubit
initial states. It is seen that they are essentially independent of the
initial state, supporting the assumption that the measurement evolution
is governed by the system-apparatus coupling alone. Using different
system-apparatus couplings and different transmon qubits, we have
obtained similar results in the range $\tau(g,t)\in[0,2]$.

\noindent{\bf Discussion:}
We note that several investigations in recent years have observed quantum
trajectories for transmons undergoing weak measurement; even the distribution
of quantum trajectories has been qualitatively presented as grey-scale
histograms in Ref.~\cite{Weber}. We have taken these observations to a
quantitative level, demonstrating that the entire trajectory distribution
can be described in terms of a single evolution parameter that is independent
of the initial qubit state and the relaxation time. (Relaxation is unavoidable
in any realistic evolution, and we accounted for it as a simple exponential
decay.) This result puts on a strong footing the quantum diffusion paradigm,
which replaces the projection postulate by an attraction towards the
measurement eigenstates plus noise, both arising from the system-apparatus
interaction with precisely related magnitudes. This unraveling of the
measurement, singling out non-universal quantum diffusion as an appropriate
extension of quantum mechanics, opens a window on to quantum physics beyond
the textbook description. Also, each quantum trajectory with its noise history
is associated with an individual experimental run, and understanding that can
help in improving quantum control and feedback mechanisms \cite{Vijay}.

Looking at quantum trajectories as physical processes has important
implications. First is the construction of new physical observables from
the distribution of quantum trajectories. The trajectories are highly
constrained, with pure states ($\rho^2=\rho$, det$(\rho)=0$) remaining
pure throughout measurement. With this restriction, any power-series
expandable function $f(\rho)$ is a linear combination of $\rho$ and $I$,
and so Tr$(f(\rho)O)$ reduces to ensemble averages that define conventional
expectation values. Defining new physical observables that characterise the
trajectory distribution is therefore a challenge.

Second, the fluctuation-dissipation relation for quantum trajectories is
a powerful clue for understanding the dynamics of quantum measurement
\cite{BornRule}. The measurement is specific to each system-apparatus pair,
with a particular value of interaction coupling and a particular type of
noise. It implies that the quantum state collapse is not universal, and the
environment can influence the measurement outcomes only via the apparatus
and not directly.

Finally, while the quantum diffusion dynamics replaces the system-dependent
Born rule by a system-independent noise, the origin of the noise remains to
be understood. That necessitates making a quantum model for the apparatus.
In our experiment, the apparatus pointer states are coherent states, with an
inherent uncertainty equal to the zero-point fluctuations. They can provide
the quantum noise for the trajectories through back-action. (The noise is
quantum and not classical, because its inclusion makes the trajectory weights
$w_0,w_1$ in Eq.(\ref{oneqweight}) go outside $[0,1]$ \cite{BornRule}.)
Understanding the irreversible quantum collapse would then amount to
understanding why sufficiently amplified coherent states do not remain
superposed \cite{Haroche}. Amplification is a driven process, with built-in
time asymmetry, and so the dynamics of the amplifier \cite{amplifrev} would
become a crucial ingredient for figuring out the quantum to classical
cross-over. This is an open field for future explorations.

\noindent{\bf Acknowledgments:}
This work was supported by the Department of Atomic Energy of the Government
of India. PK acknowledges a CSIR research fellowship from the Government of
India. RV acknowledges funding from the Department of Science and Technology
of the Government of India via the Ramanujan Fellowship.

\newpage
%\bigskip
\section*{Supplementary Information}

\subsection{Experimental Details}
The single junction 3D transmon device was fabricated using standard e-beam
lithography and double angle evaporation of aluminum on an intrinsic silicon
wafer. The device chip was placed inside a 3D aluminum cavity with a resonant
frequency $\omega_c=7.240$~GHz and a linewidth $\kappa=3.05$~MHz. The chip
was positioned away from the center of the cavity and the measured dispersive
coupling was $g_d=81.76$~MHz. The qubit frequency was measured to be
$\omega_{01}=4.93521$~GHz with an anharmonicity $\alpha=331.6$~MHz.
The dispersive shift $\chi=365$~kHz is about an order of magnitude smaller
than $\kappa$, so that the scattering phase shift and the average information
per photon become small enough to enable operation in the weak measurement
regime.

The Josephson Parametric Amplifier (JPA) is also fabricated using similar
techniques, but it is designed with a SQUID to enable tuning of the amplifier
center frequency. It is operated in the phase sensitive mode using
double-pumping technique and amplifies only the quadrature containing the
information about the $\sigma_z$ component of the qubit. The measurement
pulse, paramp pump and the demodulation signal are all generated from the
same microwave source, and variable attenuators and phase shifters are used
to control each tone.

\subsection{Data Analysis}
The observed variance of the measurement current distributions, $\sigma^2$,
is modified from its ideal value due to limited amplifier efficiency:
$\sigma^2=\sigma^2_{\rm ideal}+\sigma^2_{\rm noise}$. That affects the
trajectory reconstruction, in the sense that the actual collapse is faster
than what is seen. As a result, the fitted value of $\tau(g,t)$ automatically
includes the factor of efficiency (the actual value of $\tau$ is different).
Apart from this simple rescaling of $\tau$, limited amplifier efficiency
has no other effect on comparison between the experimental data and the
theoretical prediction. Working backwards (see \cite{ampeff}), we estimate
the amplifier efficiency to be about $0.2$ for the weaker coupling, and
about $0.35$ for the stronger coupling (for the same two data sets presented
in Figs.~\ref{eigendist},\ref{gtvst}).
% The parameters used in the Bayesian evolution formula for $t>2\mu s$ are:
% Stronger coupling: $I_0=128.92$, $I_1=127.32$, $\sigma=5.93$
% Weaker coupling: $I_0=128.44$, $I_1=127.69$, $\sigma=5.56$.

The extracted values of $I_0$ and $I_1$ showed a noticeable anomalous behaviour
in the first $2\mu s$; the former should be a constant and the latter should
decay exponentially. This anomaly is likely due to the photons left in the
cavity after the heralding/excitation pulse. To take care of it, we fitted
the observed $I_0$ and $I_1$ by exponential functions for $t>2\mu s$. Then we
constructed the evolution trajectories using the observed $I_0$ and $I_1$ for
$t\le 2\mu s$, and the fitted values of $I_0(t\rightarrow\infty)$ and
$I_1(2.5\mu s)$ for $t>2\mu s$.

To obtain $\chi^2$ values for fits of the experimental quantum trajectory
histograms with the theoretical predictions, we needed estimates of errors
for the binned histogram values. With sufficiently large trajectory ensembles,
the statistical errors are small compared to the systematic errors. The
systematic errors were not directly available, because the histograms were
produced after constructing the evolution trajectories as per
Eqs.(\ref{Bayesevol1},\ref{Bayesevol2}). So we first estimated errors in the
parameters $T_1$, initial state, $I_0$, $I_1$ that are used in construction
of the evolution trajectories. by direct analysis of the experimental data.
Then we generated different trajectory ensembles by shifting the parameter
values by their errors, one parameter at a time. Finally, assuming that
the changes in histogram values obtained for individual parameter shifts
were independent, they were added in quadrature to ascertain the total
systematic error for the binned histogram values.

Overall, we have noticed that the agreement between experiment and theory
improves with weaker couplings; we are yet to understand why. It may very
well be that the unaccounted sources of error have a larger influence at
stronger couplings.


\begin{thebibliography}{99}
\bibitem{qubitevol}{Only the diagonal elements of the density matrix
	determine the measurement outcomes, with $\rho_{11}=1-\rho_{00}$.
	Eqs.(\ref{Strat1qd},\ref{FPevol}) are simple modifications of the
	corresponding expressions in Ref.~\cite{Gisin}, to include effects
	of a finite $T_1$. We follow the notation of Ref.~\cite{BornRule}.}
\bibitem{strongtau}{For $\tau>10$, 99\% of the probability is within 1\% of
        the two fixed points \cite{BornRule}.}
\bibitem{drift}{We checked that adding a drift term to the simulated
	quantum trajectory distribution, i.e. violating the Born rule
	constraint $gS_\xi=1$, does not improve the fits.}
\end{thebibliography}
\end{document}